\documentclass[aps, prl, twocolumn]{revtex4-1}
\usepackage{graphicx}
\usepackage{float}
\usepackage{amsmath}
\usepackage{amsfonts}
\usepackage{color}

\begin{document}

\title{Bulk and surface excitons in the van der Waals magnet CrSBr: Magneto-optical studies to 55 tesla}

\author{Junho Choi$^{1,2}$, Yihyun Moon$^2$, Doohyeon Lee$^2$, Iva Plutnarov\'a$^3$, Zden\v{e}k Sofer$^3$, Vinod M. Menon$^4$, Scott A. Crooker$^{1}$}
\affiliation{$^1$National High Magnetic Field Laboratory, Los Alamos National Laboratory, Los Alamos, NM 87545, USA}
\affiliation{$^2$Department of Physics, Kyung Hee University, Seoul 02447, Republic of Korea}
\affiliation{$^3$Department of Inorganic Chemistry, University of Chemistry and Technology Prague, Prague 6, Czech Republic}
\affiliation{$^4$Department of Physics, Graduate Center of the City University of New York, New York, NY 10016, USA}

\begin{abstract}

In thin layers of the 2D magnetic semiconductor CrSBr, very recent studies identified two distinct band-edge optical resonances, believed to arise from distinguishable bulk and surface excitons. This behavior reportedly originates from the highly anisotropic nature of CrSBr --particularly in its antiferromagnetic state-- where excitons are effectively confined within individual monolayers, such that excitons in the two surface layers ``see'' a different local dielectric environment and have a lower resonance energy. To explore this scenario, here we investigate optical absorption properties of few-layer CrSBr in magnetic fields. In addition to the fundamental exciton resonance at $\approx$1.36\,eV, we observe an absorption resonance $\sim$20\,meV lower in energy. Compared to the fundamental transition, this resonance redshifts only half as much in small magnetic fields that induce ferromagnetic order, while in high fields to 55\,T it exhibits a smaller diamagnetic shift. Both behaviors point to distinguishable populations of bulk and surface excitons in CrSBr.

\end{abstract}

\maketitle

Among the emerging families of two-dimensional van der Waals (vdW) magnets \cite{Gong:2019, Gilbertini:2019, Coraux:2025, Park:2025}, the magnetic semiconductor CrSBr has focused particular interest owing to its stability in air, direct optical bandgap in the near-infrared, and variety of magnetic phases (antiferromagnetic, ferromagnetic, paramagnetic) that are readily tuned with temperature and magnetic field \cite{Telford:2020, Wilson:2021, Lee:2021, Klein:2023, Ziebel:2024, Klein:2024}.  Moreover, CrSBr couples strongly to light and hosts tightly bound electron-hole pairs (excitons) whose properties are intimately tied to the underlying magnetic order \cite{Wilson:2021, Klein:2024, Bianchi:2023, Tabataba:2024, Semina:2025, Heissenbuttel:2025, Smiertka:2025, Komar:2024}. CrSBr therefore provides a natural platform for investigating the interplay between spin, charge, and light in atomically-thin materials. 

Individual monolayers of CrSBr comprise two planes of CrS sandwiched between Br atoms (see Fig. 1a). Owing to its highly anisotropic crystalline and electronic band structure \cite{Wilson:2021, Klein:2023, Bianchi:2023, Semina:2025, Ziebel:2024, Smolenski:2025}, optically-allowed transitions at the CrSBr band-edge are linearly polarized along the in-plane $\hat{b}$ axis, and excitons have an elongated (cigar-like, nearly 1D) shape that is stretched along $\hat{b}$ but tightly localized along the in-plane $\hat{a}$ axis and the out-of-plane $\hat{c}$ axis. Crucially, excitons are largely confined within individual CrSBr monolayers due to weak inter-layer coupling, because the band-edge conduction and valence bands are composed primarily of Cr and S orbitals, with little admixture of Br \cite{Wilson:2021, Klein:2023, Ziebel:2024}. This intra-layer confinement is strongest in the A-type antiferromagnetic phase below $T_N \approx 133$\,K, wherein all Cr spins within each monolayer align ferromagnetically along $\pm \hat{b}$, while spins in neighboring monolayers orient in the antiparallel direction \cite{Goser:1990}. The opposite Cr spin orientation in neighboring layers further suppresses inter-layer hopping of band-edge electrons and holes. 

A very interesting consequence of this intra-layer confinement is that excitons residing in the two surface monolayers of a CrSBr crystal (layers 1 and $N$ in an $N$-layer sample) may be spectrally distinguishable -- and may exhibit fundamentally different properties-- in comparison to excitons confined to the internal ``bulk-like'' layers within the crystal (layers 2 through $N$-1). This is because, as depicted schematically in Fig. 1a, electrons and holes confined to the surface layers ``see'' a different local dielectric environment, which leads to a different exciton binding energy $E_b$, a locally modified free-particle bandgap $E_g$, a different exciton resonance energy ($E_g - E_b$), and a different exciton size. The strong influence of local dielectric environment on these exciton properties has been explored, both theoretically \cite{Cudazzo:2011, Komsa:2012, Latini:2015, Kylanpaa:2015, Andersen:2015, Rosner:2016, vanTuan:2024} and experimentally \cite{Lin:2014, Stier:2016, Raja:2017, Stier:2018}, in nonmagnetic semiconductor monolayers such as WSe$_2$ and MoS$_2$. In magnetic CrSBr, local dielectric effects \cite{Rudenko:2023, Piel:2026} may also be anticipated to manifest in a modified coupling of surface excitons to applied magnetic fields and the underlying magnetic order parameter, thereby providing additional avenues to experimentally test and engineer the impact of dielectric environment on 2D magnets.

A very recent study by Shao \textit{et al.} \cite{Shao:2025} provided compelling experimental evidence (and theoretical support) for \textit{distinguishable} surface and bulk excitons in CrSBr; namely, the clear identification of an additional optical resonance in few-layer samples, lying $\sim$20\,meV below the fundamental band-edge exciton, with energy and oscillator strength independent of sample thickness for $N\geq 2$. If correct, this discovery has exciting potential implications for the ability to selectively target --and spectrally address-- distinct exciton populations residing in neighboring monolayers within layered vdW magnets, enabling, for example, studies of inter-layer exciton interactions and dielectric engineering of excitonic properties. 

Here we explore --and aim to validate-- this picture of distinguishable surface and bulk excitons in CrSBr by investigating and quantitatively comparing the effects of applied magnetic fields $B$ on their optical properties, which was not done previously but for which there are clear and testable expectations.  Specifically, the well-documented coupling of excitons to the underlying magnetic order in CrSBr \cite{Wilson:2021, Ziebel:2024, Heissenbuttel:2025, Komar:2024, Smiertka:2025} is anticipated to be \textit{weaker} for surface excitons, because they have only one neighboring layer that is magnetically active, as opposed to two neighboring layers for excitons confined within internal ``bulk-like'' monolayers.  Furthermore, the characteristic average radius $r$ of a surface exciton's spatial wavefunction should be different because of different local dielectric screening \cite{Stier:2016}, which in turn should directly manifest in quantitative measurements of the exciton's quadratic diamagnetic shift, $\Delta E_{\rm dia}$, in large applied magnetic fields, because $\Delta E_{\rm dia} \propto \langle r^2 \rangle B^2$. 

In this work, we experimentally test these predictions by measuring the polarized optical absorption of few-layer CrSBr at low temperatures (4\,K) and in high magnetic fields to 55\,T. We confirm the presence of an additional absorption resonance lying $\sim$20\,meV below the fundamental band-edge exciton resonance, and find that it systematically exhibits only about half the redshift in small applied fields ($B <2$\,T) that induce ferromagnetic order. Furthermore, it systematically exhibits a smaller diamagnetic shift in high fields to 55\,T. Both of these behaviors are consistent with, and provide magneto-optical evidence in favor of, a scenario of distinguishable bulk and surface exciton populations in CrSBr.  

\begin{figure}[t] 
\center
\includegraphics[width=.95\columnwidth]{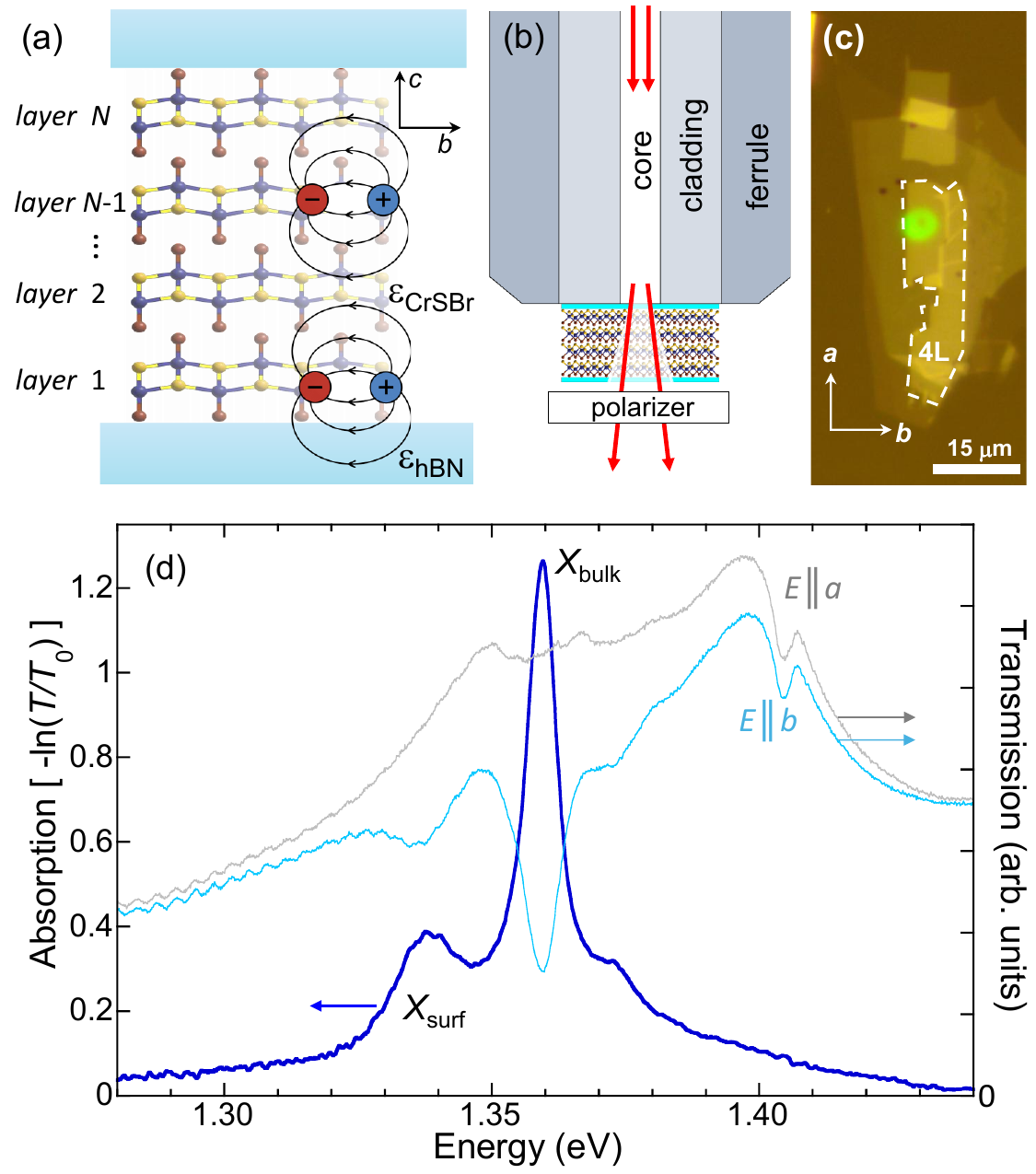}
\caption{a) Illustration of bulk and surface excitons ($X_{\rm bulk}$, $X_{\rm surf}$) in a hBN-encapsulated $N$-layer CrSBr sample. Excitons are largely confined within individual monolayers, due to weak interlayer coupling. Excitons in surface layers see a different local dielectric environment due to the neighboring hBN ($\epsilon_{\rm hBN} < \epsilon_{\rm CrSBr}$), leading to weaker screening and slightly different resonance energy. Blue, yellow, red spheres denote Cr, S, Br atoms. b) Schematic of the sample-on-fiber assembly, which enables absorbance spectroscopy of exfoliated vdW materials in high (pulsed) magnetic fields (see main text). c) Photograph of an encapsulated 4L CrSBr sample, affixed over the fiber's 5\,$\mu$m diameter core (green 532\,nm light, used for alignment, is coming through the fiber core and sample). d) The optical absorption spectrum, calculated as $-{\rm ln}(T/T_0)$, of this 4L CrSBr sample at low temperature (4\,K) and at zero magnetic field ($B$=0), using linearly-polarized light with electric field $E \parallel \hat{b}$. The reference spectrum $T_0$ was acquired using $E \parallel \hat{a}$; both raw transmission spectra $T$ and $T_0$ are also shown. The absorption resonances from $X_{\rm bulk}$ and $X_{\rm surf}$ have Lorentzian lineshapes and are centered at $\approx$1.359\,eV and 1.338\,eV, respectively, in this 4L sample.
}
\label{fig1}
\end{figure}

We focus on measurements of optical absorption spectroscopy, because it provides an especially direct measure of absorption oscillator strengths in materials, in comparison to the more commonly-utilized reflectivity spectroscopy for which Kramers-Kronig analysis and transfer-matrix formalism are generally required to account for multi-layer interference effects from thin layered materials \cite{Shao:2025, Raja:2019, Byrnes:2016}. To measure the optical absorption of small, few-layer exfoliated CrSBr samples in high (pulsed) magnetic fields, we adopt a sample-on-fiber approach that has previously enabled absorption studies of non-magnetic monolayer semiconductors \cite{Stier:2018, Goryca:2019, Li:2020}.  Figure 1b shows a schematic of the assembly, where an exfoliated CrSBr flake is encapsulated between thin slabs of hexagonal boron nitride (hBN), and then placed over the core of a single-mode optical fiber using standard methods for vdW dry stacking (the CrSBr thickness was measured separately by atomic force microscopy). Broadband white light from a xenon lamp was then directed through the fiber and the sample.  Thin-film linear or circular polarizers were used to select out specific optical polarizations, and the transmitted light was then retro-reflected back into a separate multimode collection fiber and then dispersed in a spectrometer and detected by a cooled CCD camera. A photomicrograph of an encapsulated CrSBr sample, affixed over a fiber core, is shown in Fig. 1c.  The sample assembly was mounted on fiber-coupled probe and loaded into a helium bath cryostat, in the bore of a capacitor-driven pulsed magnet. This sample-on-fiber approach ensures a rigid and robust optical path through the sample, that is unaffected by the vibrations and drift that can exist in low-temperature measurements in magnetic fields.  

Figure 1d shows the optical absorption spectrum, calculated as $-{\rm ln}(T/T_0$), of a 4-layer CrSBr sample at low temperature (4\,K) and zero field ($B$=0).  To best capture the exciton absorption, the transmission spectrum $T$ and the reference spectrum $T_0$ were acquired using light linearly polarized along the crystal's $\hat{b}$ and $\hat{a}$ axis (where band-edge exciton absorption in CrSBr is maximal, and where it vanishes, respectively). In this way polarization-independent spectral features cancel out, as seen in the Fig. 1d.  The strong and narrow resonance from the well-studied fundamental band-edge exciton at $\approx$1.359\,eV dominates the spectrum.  A weak high-energy shoulder at $\approx$1.373\,eV is also present; this commonly-observed spectral feature \cite{Wilson:2021, Klein:2023, Tabataba:2024, Semina:2025, Shao:2025, Krelle:2025} is often ascribed to the nominally forbidden optical transitions from upper conduction bands.  Most importantly, a prominent additional absorption resonance is also plainly revealed at $\approx$1.338\,eV, which is $\sim$21\,meV \textit{below} the fundamental exciton resonance.  

This additional lower-energy resonance is in close correspondence with the recent results of Shao \textit{et al.} \cite{Shao:2025}, who observed a lower-energy optical resonance in reflectivity studies of few-layer (and also bulk) CrSBr.  Based on their finding that its energy and oscillator strength did not vary with sample thickness, and by also considering its evolution with temperature, Shao \textit{et al.} proposed that the lower-energy resonance originates from  excitons confined in the two surface monolayers of their CrSBr samples. Such surface excitons necessarily see a different (typically less-screened) dielectric environment in comparison to excitons confined to the bulk-like monolayers within the crystal, and the concomitant reductions of the free-particle bandgap $E_g$ and the exciton binding energy $E_b$ do not exactly compensate, leading to an optical transition energy ($E_g - E_b$) that is slightly less than that of the more highly screened bulk-like excitons within the CrSBr crystal. 

The proposed picture of distinguishable populations of bulk and surface excitons in CrSBr invites (at least) two experimental tests in applied magnetic fields $B$, which to date have not yet been explored: First, it is well established that excitons in CrSBr are intimately coupled to the underlying magnetic order, such that the 1.36\,eV band-edge exciton resonance redshifts by $\approx$15\,meV when CrSBr is driven from antiferromagnetic to ferromagnetic order by small $B$ \cite{Wilson:2021, Tabataba:2024, Heissenbuttel:2025, Ziebel:2024, Komar:2024, Smiertka:2025}. For $B$ applied along the hard (out-of-plane, $\hat{c}$) magnetic axis, the redshift is parabolic in $B$ as the Cr spins smoothly cant towards $\hat{c}$ until out-of-plane ferromagnetic alignment of all the layers is achieved for $|B| \gtrsim 2$\,T. The redshift in the ferromagnetic state arises because interlayer hopping of electrons and holes is no longer quite as strongly suppressed (by the antiferromagnetic spin orientation in neighboring layers), and the exciton's spatial wavefunction can expand (delocalize) slightly in both the $+\hat{c}$ and $-\hat{c}$ directions, such that it is no longer as strictly confined to a single CrSBr monolayer. This spatial deconfinement leads to a reduction of the exciton's transition energy. Crucially, however, for surface excitons, this deconfinement can necessarily occur only in one $\hat{c}$ direction, not both, and a smaller redshift can be anticipated.  

\begin{figure}[t]
\center
\includegraphics[width=.95\columnwidth]{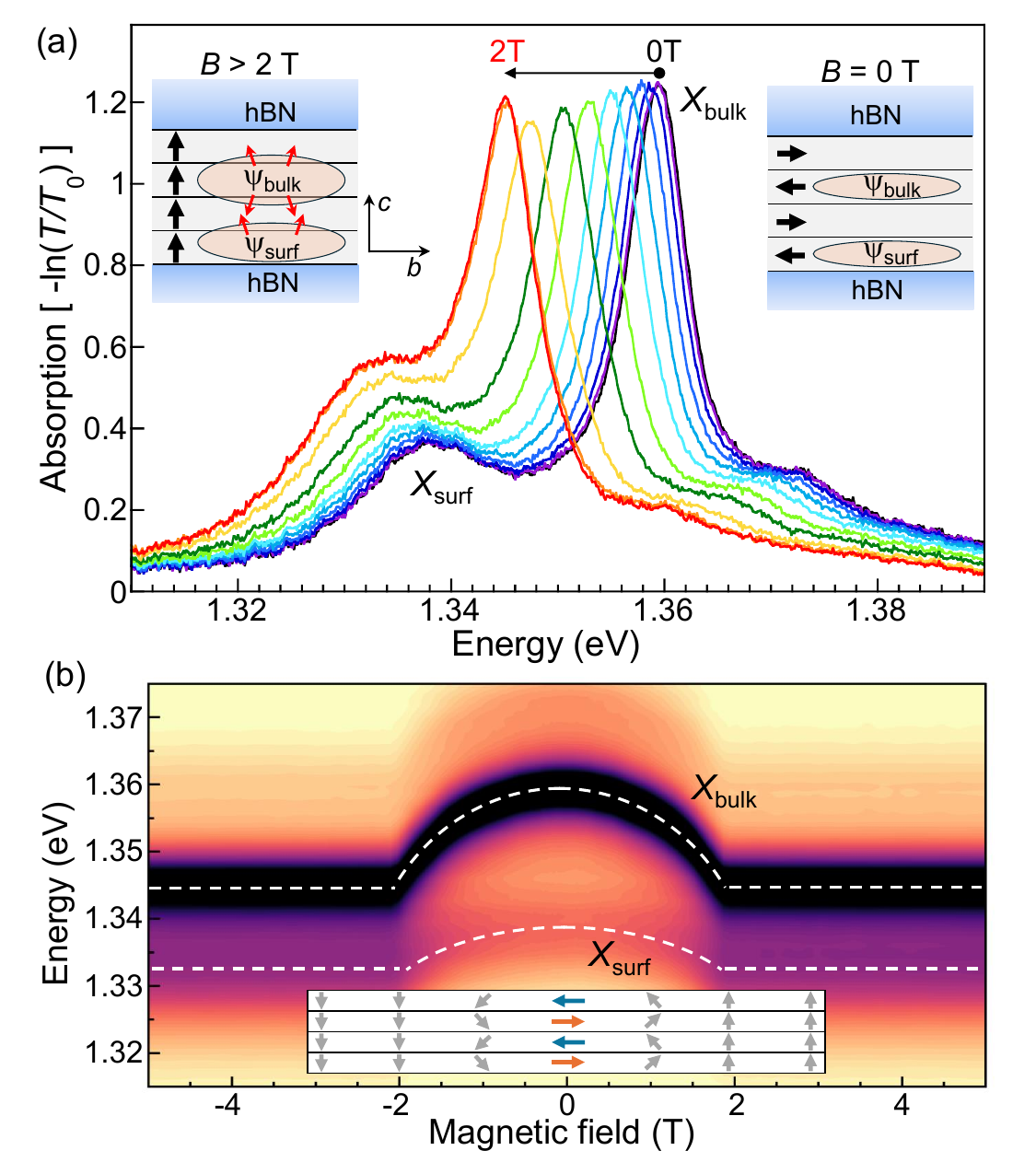}
\caption{a) Low-temperature absorption spectra from 4L CrSBr with increasing $B~(\parallel \hat{c})$ from $0 \rightarrow 2$\,T (in steps of $\sim$0.2\,T). Both $X_{\rm bulk}$ and $X_{\rm surf}$ resonances redshift to lower energies as the CrSBr reorients from antiferromagnetic to ferromagnetic order. The redshift of $X_{\rm surf}$ ($\sim$6\,meV) is slightly less than half that of $X_{\rm bulk}$ ($\sim$14\,meV). The red ovals in the cartoon insets represent their spatial envelope wavefunctions $\psi_{\rm surf}(r)$ and $\psi_{\rm bulk}(r)$ within the CrSBr sample, and  depict how $\psi_{\rm bulk}(r)$ delocalizes in both $+\hat{c}$ and $-\hat{c}$ directions when interlayer coupling becomes more likely in the ferromagnetic state, whereas $\psi_{\rm surf}(r)$ can delocalize in only the $+\hat{c}$ direction (or $-\hat{c}$ for the top surface), leading to its smaller redshift. b) A 2D plot of the energy- and $B$-dependent absorption, showing more clearly the large and small parabolic redshifts of $X_{\rm bulk}$ and $X_{\rm surf}$ .}
\label{fig2}
\end{figure}

Secondly, it is also well known from studies of excitons in monolayer semiconductors \cite{Stier:2016, Stier:2018} that local dielectric environments with weaker screening (smaller $\epsilon$) result in excitons with spatial wavefunctions that are physically smaller, which leads to smaller measured diamagnetic shifts in applied $B$. We recall that diamagnetic shifts are proportional to the square of the exciton's characteristic radius $r_\perp$ in the plane perpendicular to $B$,

\begin{equation}
\Delta E_{\mathrm{dia}} = \frac{e^{2}}{8 \mu} \langle r_\perp^{2} \rangle B^{2} = \sigma B^{2},
\end{equation}

where $\langle r_\perp^2 \rangle = \langle \psi | x^2 + y^2 | \psi \rangle$ is the expectation value of $r_\perp^2$ calculated over the exciton's spatial wavefunction $\psi(r)$, $\mu$ is the exciton's reduced mass, and $\sigma$ is the diamagnetic shift coefficient. Equation (1) remains valid for arbitrarily-shaped wavefunctions $\psi(r)$, and applies in the ``weak-field'' limit where the magnetic length $l_B = \sqrt{\hbar/eB}$ ($\approx$26/$\sqrt{B}$ nm) exceeds the exciton radius, which is the case for CrSBr and the fields used in these studies.  Surface and bulk excitons should therefore exhibit different diamagnetic shifts.  Specifically, because the dielectric constant of the hBN encapsulating layers ($\epsilon_{\rm hBN} \approx 4.5$) is smaller than that of CrSBr (estimates and measurements for $\epsilon_{\rm CrSBr}$ lie in the range of $\approx$10 \cite{Dirnberger:2023, Piel:2026}), surface excitons experience less dielectric screening, are expected to have smaller lateral extent within the $a-b$ plane, and should exhibit smaller diamagnetic shifts.

In Figure 2 we explore the first test discussed above, and show how the exciton resonances redshift in small $B \parallel \hat{c}$, which drives the CrSBr from its antiferromagnetic to ferromagnetic phase. The fundamental absorption redshifts by $\approx$14\,meV (from 1.3594\,eV to 1.3452\,eV) as the Cr spins in the sample cant and reorient from A-type antiferromagnetism to a ferromagnetic state where all layers are magnetized along $\hat{c}$ by 2\,T. The parabolic nature of the redshift is seen clearly in the 2D map shown in Fig. 2b, and has been observed many times in recent studies of CrSBr \cite{Wilson:2021, Tabataba:2024, Heissenbuttel:2025, Klein:2024}. More importantly, Fig. 2 also shows that the lower-energy exciton resonance also redshifts, but only by about 6\,meV (from 1.3380\,eV to 1.3324\,eV).  As discussed above, and as the cartoons in the inset of Fig. 2a depict, this is consistent with the assignment of the lower-energy resonance to surface excitons, which can delocalize only in one $\hat{c}$ direction as interlayer hopping becomes slightly more allowed in the ferromagnetic state, in contrast to excitons confined within bulk-like layers in CrSBr which can delocalize in \textit{both} $\pm \hat{c}$ directions (thus approximately doubling the redshift, if the degree of delocalization is small). We emphasize that, as recent \textit{ab initio} calculations by Heissenb\"uttel \textit{et al.} have shown \cite{Heissenbuttel:2025}, the spatial wavefunction of optically-allowed excitons delocalizes only by a limited amount along $\hat{c}$ (acquiring only $\sim$10\% inter-layer character), so that the notion of layer-confined excitons still applies even in the ferromagnetic state. As discussed later in Fig. 4, these overall trends are also consistently observed in 2-layer, 3-layer, and in thick CrSBr samples.

\begin{figure}
\center
\includegraphics[width=.95\columnwidth]{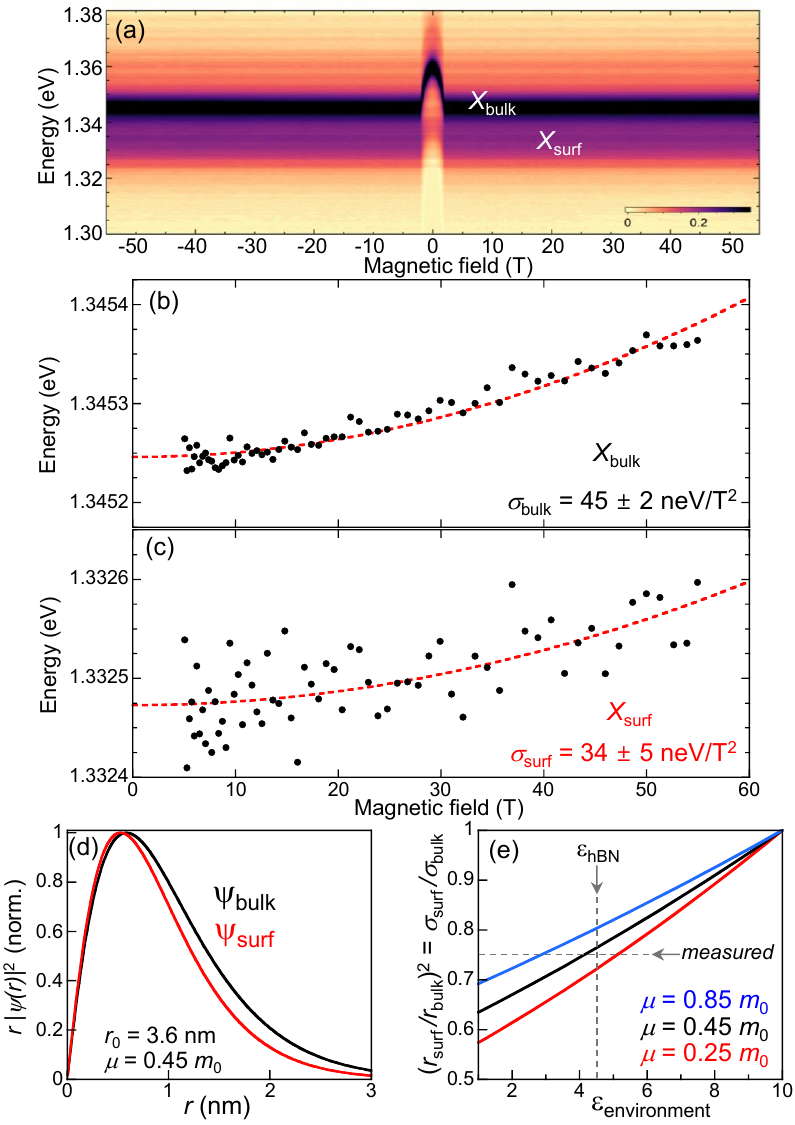}
\caption{a) Absorption of the 4L CrSBr sample in high magnetic fields to $\pm$55\,T, using circularly polarized light. b,c) The quadratic diamagnetic shift of the bulk and surface exciton resonances. The diamagnetic shift of $X_{\rm surf}$ is 25\% less than that exhibited by $X_{\rm bulk}$, consistent with the reduced dielectric screening of $X_{\rm surf}$ due to the hBN encapsulation. d) Calculated radial probability density of bulk and surface exciton wavefunctions in hBN-encapsulated CrSBr (see main text). e) Calculated ratio $\sigma_{\rm surf} / \sigma_{\rm bulk}$ for CrSBr, versus dielectric constant of the surrounding environment, for different exciton reduced masses $\mu$.}
\label{fig3}
\end{figure}

In Figure 3 we investigate the second conjecture, and measure the diamagnetic shifts of these exciton resonances in high magnetic fields to $\pm 55$\,T. Here we detect circularly-polarized light, because field-induced polarization (Faraday) rotation of light in optical components, combined with linear polarizers, can lead to intensity modulation artifacts in the detected signals. The entire field dependence of the absorption spectra is shown in the color map of Fig. 3a. For all $|B| \gtrsim 2$\,T (that is, in the ferromagnetic state of CrSBr), the detailed shape of the absorption spectra are essentially unchanged except for the small diamagnetic shifts which are on the order of 100\,$\mu$eV. However, the good signal-to-noise of the measured absorption signals (\textit{cf} Fig. 2a) allows to accurately fit these resonances to Lorentzian oscillators.  Figures 3b and 3c show the field-dependent shifts exhibited by the two exciton resonances. Both follow a quadratic $B^2$ dependence, consistent with diamagnetic  shifts, with coefficients $\sigma = 45 \pm 2$\,neV/T$^2$, and $\sigma = 34 \pm 5$\,neV/T$^2$.  Crucially, the diamagnetic shift of the lower-energy absorption resonance is $\sim$25\% smaller, indicating that it originates from an exciton state with 25\% smaller characteristic mean-squared radius $\langle r_\perp^2 \rangle$, which is consistent with a picture of excitons in CrSBr surface layers that experience weaker local dielectric screening in comparison to those in bulk-like layers. Along with the low-field exciton redshifts shown in Fig. 2, the high-field diamagnetic shifts shown in Fig. 3 provide the second key new piece of magneto-optical evidence supporting the proposed scenario \cite{Shao:2025} in which surface excitons in CrSBr generate a distinguishable optical resonance at lower energy.

\begin{figure*}
\center
\includegraphics[width=1.8\columnwidth]{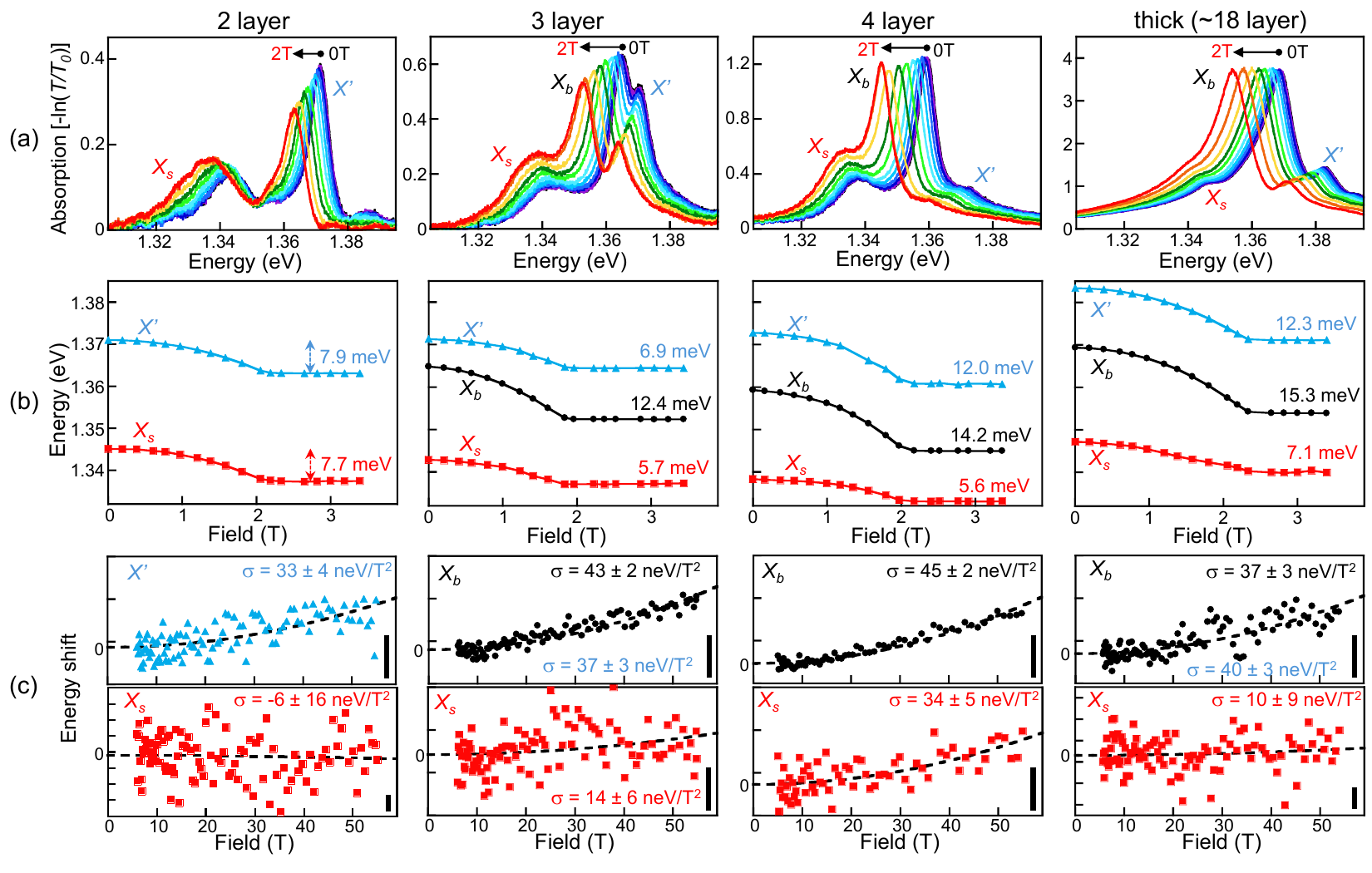}
\caption{a) The four panels show low-temperature optical absorption spectra from a 2L, 3L, 4L, and a thick ($\sim$18L) CrSBr sample respectively, as $B \parallel \hat{c}$ increases from $0 \rightarrow 2$\,T (in steps of $\approx$0.2\,T). Note that the 2L sample lacks the central absorption resonance from $X_{\rm bulk}$, because no ``bulk-like'' layers exist in 2L CrSBr. b) These panels show the corresponding $B$-dependent energies of the absorption resonances. The net redshifts of the resonances (between 0\,T and $\sim$2\,T) are indicated. In all cases, the redshift of $X_{\rm surf}$ is slightly less than half the redshift of $X_{\rm bulk}$. c) These panels show the high-field diamagnetic shifts of these absorption resonances, along with best quadratic fits.  Tick marks on the vertical axes indicate 0.1\,meV intervals, and are also represented by the vertical scale bars.  The diamagnetic coefficients $\sigma$ are indicated. The spectra and the diamagnetic shifts for the 4L sample are the same as shown in previous Figures, but are reproduced here for completeness and comparison. To within measurement uncertainty, diamagnetic shifts of $X_{\rm surf}$ are consistently less than those of $X_{\rm bulk}$.}
\label{fig4}
\end{figure*}

In an idealized material hosting excitons with isotropic reduced mass $\mu = 1 m_0$, the measured diamagnetic coefficient $\sigma$=45\,neV/T$^2$ would indicate small excitons with $r_\perp$=1.45\,nm. However, CrSBr is highly anisotropic, with both theoretical and experimental studies \cite{Wilson:2021, Klein:2024, Smiertka:2025, Heissenbuttel:2025, Smolenski:2025} indicating anisotropic exciton masses ($\mu_b \approx 0.1\,m_0$, $\mu_a \approx 2\,m_0$) and elongated excitons that extend $\approx$3\,nm and 1\,nm along the $\hat{b}$ and $\hat{a}$ axes, respectively. Such tightly-bound band-edge excitons are on the border between ``Wannier-like'' and ``Frenkel-like'' \cite{Semina:2025, Smiertka:2025}, and an accurate comparison of the measured diamagnetic shifts requires a suitable integration over the calculated exciton wavefunctions. Nonetheless, we can quantitatively estimate exciton sizes and diamagnetic shift ratios assuming isotropic excitons with reduced mass $\mu = \sqrt{\mu_a \mu_b} \approx 0.45\,m_0$ (the geometric mean of in-plane masses), within the framework of the Rytova-Keldysh potential for a 2D material sandwiched above and below by other dielectrics \cite{Kylanpaa:2015, Stier:2016}: $V_{RK}(r) = \frac{-e^2}{8\epsilon_0 r_0}[H_0(\frac{\kappa r}{r_0}) - Y_0(\frac{\kappa r}{r_0})]$. Here, $H_0$ and $Y_0$ are Struve and Bessel functions of the second kind, $\kappa = \frac{1}{2}(\epsilon_{\rm above} + \epsilon_{\rm below})$ is the average dielectric constant of the surrounding materials, and $r_0 = \frac{d}{2}(\epsilon_{\rm CrSBr} - 1) = 3.6$\,nm is the characteristic screening length of a CrSBr monolayer (thickness $d$=0.8\,nm). 

Figure 3d shows the calculated radial wavefunctions $\psi(r)$ for excitons in bulk monolayers where $\kappa = \epsilon_{\rm CrSBr}$, and for (the smaller) excitons in surface monolayers where $\kappa = \frac{1}{2}(\epsilon_{\rm CrSBr} + \epsilon_{\rm hBN}) = \frac{1}{2}(10 + 4.5) = 7.25$.  Figure 3e shows the calculated ratio $\sigma_{\rm surf} / \sigma_{\rm bulk}$ (= $r^2_{\rm surf}/r^2_{\rm bulk}$) versus the dielectric constant of the surrounding environment, for different masses $\mu$. The calculated ratio is in quite good agreement with the measured ratio of 0.75, using $\mu = 0.45\,m_0$. To summarize, the distinct diamagnetic shifts point to two distinguishable species of excitons in CrSBr: well-screened excitons residing in the bulk layers, and less-screened excitons confined to the surface layers.

These overall magneto-optical trends in both low and high $B$ are consistent across a range of CrSBr samples. Figure 4 shows results from 2-layer (2L), 3-layer (3L), and thick ($\sim$18L) CrSBr samples.  Absorption spectra from $0 \rightarrow 2$\,T are shown in Fig. 4a, the energies and low-field redshifts of the individual absorption resonances are shown in Fig. 2b, and the high-field (diamagnetic) shifts are shown in Fig. 4c. Crucially, all samples \textit{except} for the 2L sample exhibit three absorption resonances, which are assigned to $X_{\rm surf}$, $X_{\rm bulk}$, and the highest-energy resonance (denoted  $X'$) that may arise from nominally-forbidden transitions to an upper conduction band. The 2L sample lacks the resonance associated with $X_{\rm bulk}$, in line with the expectation that 2L CrSBr has only surface layers and no bulk layers. We further note that these data are in quite good agreement with the exciton resonances recently tracked in reflection studies of thin CrSBr by Tabataba-Vakili \cite{Tabataba:2024}, although the assignments of the individual peaks differ.

Figure 4b confirms that the lowest-energy resonance consistently exhibits slightly less than half the low-field parabolic redshift in comparison to the most prominent resonance that appears when $N \geq 3$ (i.e., 5-7\,meV \textit{vs}. 12-15\,meV), supporting their assignments as $X_{\rm surf}$ and $X_{\rm bulk}$, respectively, as discussed above. 

\begin{figure}
\center
\includegraphics[width=.8\columnwidth]{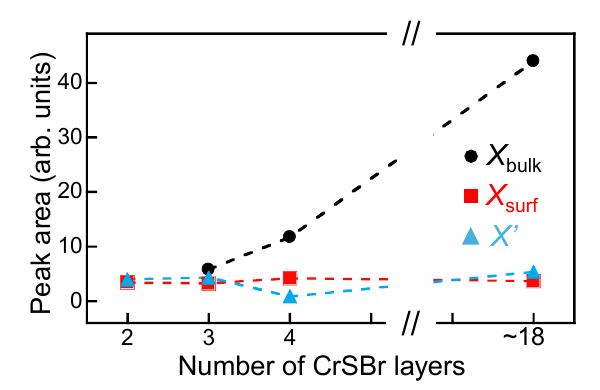}
\caption{The oscillator strengths (i.e., the area of the absorption resonances) of all excitons are shown as a function of CrSBr layer thickness $N$. The oscillator strength of $X_{\rm bulk}$ increases with thickness for $N \geq 3$, while that of $X_{\rm surf}$ remains small and approximately constant, in agreement with the expectation that multilayer CrSBr samples have only two surface layers.}
\label{fig3}
\end{figure}

Lastly, Fig. 4c shows that $\sigma_{\rm bulk}$ in the 3L and thick samples are $\approx$43 and 37 neV/T$^2$, respectively, in line with that from the 4L CrSBr sample discussed earlier, and also in approximate agreement with the high-field diamagnetic shift of the reflection resonance from bulk CrSBr reported by Smiertka et al. \cite{Smiertka:2025}. More importantly, these data also show that, for every sample we studied, $\sigma_{\rm surf}$ is consistently smaller than $\sigma_{\rm bulk}$ to within experimental uncertainty, again in support of a picture where $X_{\rm surf}$ experiences weaker dielectric screening and are physically smaller than $X_{\rm bulk}$. Where such fits were possible, the diamagnetic shifts of the highest-energy ($X'$) resonance are also shown or indicated in the panels in blue colors; they are comparable to those of $X_{\rm bulk}$. 

The extracted absorption oscillator strengths of all the exciton resonances are summarized in Fig. 5, for all measured CrSBr sample thicknesses.  Crucially, the net oscillator strength of $X_{\rm bulk}$ appears only when $N$=3 and doubles when $N$=4 (to within 10\%, which is the estimated uncertainty of these extracted values), while that of $X_{\rm surf}$ remains approximately constant (to within $\pm 15$\%), in agreement with the expectation that any multilayer CrSBr sample has $N-2$ bulk layers but only two surface layers. (The extracted oscillator strength of $X'$ is also shown; it remains small across all thicknesses and shows large sample-to-sample variations suggesting that it may arise from surface inhomogeneities or defects.)

Taken all together, the low-field redshifts and high-field (diamagnetic) blueshifts of the different exciton resonances provide new magneto-optical evidence consistent with a picture of distinguishable species of bulk and surface excitons in the vdW magnet CrSBr.  Being spectrally distinct, whilst originating from proximal atomic layers (for example in 3L or 4L CrSBr), strongly motivates future optical studies of inter-species exciton interactions and distinguishable exciton-magnon dynamics \cite{Bae:2022, Li:2024, Datta:2025}, and opens the door to dielectric engineering and control of these exciton energies. 

\textbf{Acknowledgments}: We thank M. Semina, M. Glazov, and M. Goryca for helpful discussions, and we acknowledge support from the US Department of Energy (DOE) ``Science of 100\,T'' program. The National High Magnetic Field Lab is supported by the National Science Foundation DMR-2128556, the State of Florida, and the US DOE. This work was also supported by the National Research Foundation of Korea grant funded by the Korean government (MSIP) (RS-2025-25466315 and RS-2026-25497294). Work at City College was primarily supported by the U.S. DOE Office of Science, Basic Energy Sciences (BES), under Award DE-SC0025302. Z.S. was supported by project LUAUS25268 from Ministry of Education Youth and Sports (MEYS) and by the project Advanced Functional Nanorobots (reg. No. CZ.02.1.01/0.0/0.0/15 003/0000444 financed by the EFRR.


\end{document}